%% file: main.tex
\documentclass[submission,copyright,creativecommons]{eptcs}
\providecommand{\event}{HCVS 2025} 

\usepackage{iftex}

\ifpdf
  \usepackage{underscore}         
  \usepackage[T1]{fontenc}        
\else
  \usepackage{breakurl}           
\fi

\usepackage{graphicx}
\usepackage{orcidlink}
\usepackage{todonotes}
\usepackage{amsmath,amssymb}
\usepackage{listings}
\usepackage{tikz}
\usepackage{multirow}
\usepackage{collcell}
\usepackage{booktabs}
\usepackage{float}
\usepackage{enumitem}
\usepackage{rotating}
\usepackage{subcaption}
\usepackage{numprint}
\npthousandsep{\,}
\usepackage{array}
\newcommand{\PreserveBackslash}[1]{\let\temp=\\#1\let\\=\temp}
\newcolumntype{C}[1]{>{\PreserveBackslash\centering}p{#1}}
\newcolumntype{R}[1]{>{\PreserveBackslash\raggedleft}p{#1}}
\newcolumntype{L}[1]{>{\PreserveBackslash\raggedright}p{#1}}
\usepackage{doi}

\usetikzlibrary{shapes.geometric}
\usetikzlibrary{positioning}
\usetikzlibrary{arrows}
\usetikzlibrary{shapes.multipart}
\usetikzlibrary{fit,calc}
\usepackage{pgfplots}
\usepackage{pgfplotstable}
\pgfplotsset{compat=1.18}

\newcommand*\rot{\rotatebox{90}}

\makeatletter
\let\orgdescriptionlabel\descriptionlabel
\renewcommand*{\descriptionlabel}[1]{%
  \let\orglabel\label
  \let\label\@gobble
  \phantomsection
  \edef\@currentlabel{#1\unskip}%
  \let\label\orglabel
  \orgdescriptionlabel{#1}%
}
\makeatother

\title{Theta as a Horn Solver}

\author{Levente Bajczi\orcidlink{0000-0002-6551-5860} \hspace{5ex} Milán Mondok\orcidlink{0000-0001-5396-2172} \hspace{5ex} Vince Molnár\orcidlink{0000-0002-8204-7595}
\institute{Department of Artificial Intelligence and Systems Engineering\\Faculty of Electrical Engineering and Informatics\\Budapest University of Technology and Economics\\Műegyetem rkp. 3., H-1111 Budapest, Hungary.\\\email{\{bajczi,mondok,molnarv\}@mit.bme.hu}}}

\newcommand{\titlerunning}{Theta as a Horn Solver}
\newcommand{\authorrunning}{Bajczi et al.}

\hypersetup{
  bookmarksnumbered,
  pdftitle    = {\titlerunning},
  pdfauthor   = {\authorrunning},
  pdfsubject  = {HCVS'25},
  pdfkeywords = {CHC, BMC, KIND, MDD, saturation, verification} 
}

\definecolor{ftsrg@AccentBlue}{RGB}{20,70,160}
\definecolor{ftsrg@AccentRed}{RGB}{150,0,24}
\definecolor{ftsrg@AccentPurple}{RGB}{82,43,71}
\definecolor{ftsrg@AccentOrange}{RGB}{251,139,36}
\definecolor{ftsrg@AccentLightBlue}{RGB}{68,114,196}
\definecolor{ftsrg@AccentGreen}{RGB}{112,173,71}

\begin{document}

\maketitle

\begin{abstract}
\textsc{Theta} is a verification framework that has participated in the CHC-COMP competition since 2023. While its core approach -- based on transforming constrained Horn clauses (CHCs) into control-flow automata (CFAs) for analysis -- has remained mostly unchanged, \textsc{Theta}’s verification techniques, design trade-offs, and limitations have remained mostly unexplored in the context of CHCs. This paper fills that gap: we provide a detailed description of the algorithms employed by Theta, highlighting the unique features that distinguish it from other CHC solvers. We also analyze the strengths and weaknesses of the tool in the context of CHC-COMP benchmarks. Notably, in the 2025 edition of the competition, Theta’s performance was impacted by a configuration issue, leading to suboptimal results. To provide a clearer picture of Theta’s actual capabilities, we re-execute the tool on the competition benchmarks under corrected settings and report on the resulting performance.
\end{abstract}

\begin{footnotesize}
		\paragraph{\footnotesize{Funding.}}
		This research was partially funded by the 2024-2.1.1-EKOP-2024-00003 University Research Scholarship Programme under project numbers EKOP-24-3-BME-\{78,213\}, and the Doctoral Excellence Fellowship Programme under project number 400434/2023; with the support provided by the Ministry of Culture and Innovation of Hungary from the NRDI Fund.
\end{footnotesize}

\section{Introduction}
\label{Introduction}

Constrained Horn clauses (CHCs) have emerged as a standard intermediate representation for various verification tasks (such as program- or smart-contract verification \cite{programverif,smartcontractverif}) and are the focus of the annual CHC-COMP competition\cite{chccomp}. Over the past three years, the \emph{Theta} framework\footnote{\url{https://github.com/ftsrg/theta}}, an open-source, modular model checker developed at the Critical Systems Research Group, Budapest University of Technology and Economics \cite{theta-fmcad2017,thetajar} has participated as a CHC solver in CHC-COMP, aiming to adapt techniques from software model checking to solve constrained horn clauses.

Originally conceived as a framework for predicate abstraction with configurable refinement strategies, Theta’s CHC-solving capabilities are built upon a pre-analysis transformation that converts CHC systems into control-flow automata (CFAs), enabling the reuse of mature abstraction-based algorithms, and more novel, bounded techniques~\cite{emergentheta}. This transformation step has been described in an earlier publication~\cite{bottomsup}, but beyond that, there has been no comprehensive documentation of how Theta handles CHC inputs, nor an in-depth assessment of its effectiveness across different categories of CHC problems.

This paper fills that gap. We describe in detail the architecture of Theta, and its main algorithms -- including abstraction-refinement and bounded techniques -- including the sequential portfolio we use for automated verification, with next to no need for user input when choosing a performant algorithm. We also analyze how well these methods scale and where they fall short, both in theory and in practice.

We also reflect on Theta’s 2025 CHC-COMP submission, whose performance was diminished by a misconfiguration that resulted in the first step of the portfolio never advancing to others. While the competition results for this year are already finalized, they do not accurately represent the tool's capabilities. To address this, we rerun Theta on the official benchmark set under corrected settings and present a comparative analysis of the results.

The remainder of the paper is structured as follows. In Section~\ref{approach}, we describe Theta’s CHC-solving approach in detail. Section~\ref{architecture} presents the overall software architecture of the tool, including its modular design and extensibility. Section~\ref{strengths} evaluates the key strengths and limitations of Theta's approach, informed by our experience in CHC-COMP and beyond. Section~\ref{toolsetup} outlines the specific configuration and setup used in the 2025 CHC-COMP submission, highlighting the cause of the performance issues and detailing our re-evaluation methodology. Finally, Section~\ref{softproj} provides information on the availability of the Theta tool, its source code, and the data used in our experiments to support reproducibility.

\section{Verification Approach}
\label{approach}

Theta supports multiple verification techniques for analyzing software systems encoded as Constrained Horn Clauses (CHCs). These techniques are grounded in symbolic model checking and operate over control flow automata (CFA), which are derived via a structural transformation from the original CHC system~\cite{bottomsup}. This section outlines the main verification approaches implemented in Theta, with a focus on the ones used in the CHC-COMP configuration.

\subsection{CHC-to-CFA Transformation}

As detailed in prior publications~\cite{bottomsup,chc2c}, a CHC problem can be transformed to a program-like structure -- namely, a control flow automaton (CFA) \cite{thetajar} with variables $V = \{v_1, v_2, \dots, v_n\}$ over domains $D_{v_1}, D_{v_2}, \dots, D_{v_n}$, locations $L$, and edges $E \subseteq L \times \mathit{Ops} \times L$, where $\mathit{Ops}$ can have:
\begin{itemize}[nosep,noitemsep]
    \item $\mathit{assume}(\mathit{expr})$,
    \item $\mathit{assign}(\mathit{expr}_\mathit{lhs}, \mathit{expr}_\mathit{rhs})$, and
    \item $\mathit{havoc}(v \in V)$ 
\end{itemize}
instructions for guards, variable updates, and nondeterministic value assignments, respectively. Domains of variables are subsets of domains in SMT.

Multi-procedure programs can be represented as a set of named CFAs, where procedures can be called as part of an expression. We assume all procedures are side-effect free and behave like mathematical functions (a sane assumptions, given they were transformed from the uninterpreted predicates of a CHC problem).

With a \emph{forward}, or \emph{bottom-up} transformation~\cite{bottomsup}, the resulting artifact is always represented as a single CFA, i.e., only only a single procedure will exist, and no function invocations exist in the expressions.

With a \emph{backward}, or \emph{top-down} transformation~\cite{bottomsup}, the result may be multiple CFAs, each invoking zero or more other CFA procedures as functions. In the case of non-linear clauses, only this method can be used.

\subsection{CEGAR-based Verification}

\input{figures/cegarloop}

Theta's core verification engine is based on the Counterexample-Guided Abstraction Refinement (CEGAR) paradigm~\cite{Clarke2003}, where verification is performed by iteratively refining an abstract model of the system. The main data structure maintained during this process is the \emph{Abstract Reachability Graph} (ARG)~\cite{beyer2007blast}, which over-approximates the concrete reachable state space.

In the ARG, abstract states represent sets of concrete states. This abstraction ensures soundness: if a concrete error state is reachable, then so is an abstract error state. However, the converse is not guaranteed, leading to potential \emph{spurious counterexamples}. Figure~\ref{fig:arg} illustrates this: an error state may appear reachable within the abstract graph, but it may not correspond to a concrete execution path, prompting refinement.

A key mechanism in ARG-based abstraction is \emph{state covering}. If an abstract state $s_1$ is logically implied by another state $s_2$ -- for example, if $s_1$ includes constraints $a = 2 \wedge b = 3$ and $s_2$ includes only $a = 2$ -- then $s_1$ is \emph{covered} by $s_2$. This allows the search to avoid exploring redundant paths, improving scalability. This is particularly useful in programs with loops that do not influence the error condition: intermediate states can be covered by the loop header, avoiding repeated exploration.

The CEGAR process operates in a loop, as shown in Figure~\ref{fig:cegarloop}. The loop consists of two main components: the \emph{abstractor} and the \emph{refiner}. The abstractor constructs or expands the ARG based on the current abstraction precision. It checks whether a (possibly spurious) error state is reachable. If no such state is reachable, the system is proven \emph{safe}, since the abstraction is conservative.

If an error state is found, the abstractor extracts one or more \emph{abstract counterexamples}-paths in the ARG leading to the error. These are handed to the refiner, which first checks whether the path is feasible in the concrete system. If a feasible path is found, the program is \emph{unsafe}. Otherwise, the refinement process strengthens the abstraction by computing a more precise abstraction (typically by deriving new predicates) and pruning the ARG to the point of infeasibility. The refined abstraction is then passed back to the abstractor, and the cycle continues.

The CEGAR loop described so far is a high-level specification focusing on outcomes rather than implementation details. This reflects CEGAR’s inherent modularity: its components can be freely interchanged as long as they fulfill compatible roles.

\textsc{Theta}’s CEGAR implementation emphasizes modularity, offering multiple configurable components. Among these, the two most critical are the \emph{abstract domain} and the \emph{refinement algorithm} (Implementation details not directly relevant to this paper are documented in \cite{thetajar}).

\subsubsection{Abstract Domain}

The abstract domain defines the abstraction’s foundation. \textsc{Theta} supports two main domains: the \emph{explicit value} domain and the \emph{predicate} domain. The predicate domain includes variants such as cartesian, boolean, and split boolean abstractions.

Formally, an abstract domain is a tuple \(D = (S, \top, \bot, \sqsubseteq, expr)\), where:
\begin{itemize}[noitemsep,nosep]
    \item \(S\) is a lattice of abstract states,
    \item \(\top \in S\) is the top (most abstract) element,
    \item \(\bot \in S\) is the bottom (contradictory) element,
    \item \(\sqsubseteq\) is the partial order on \(S\),
    \item \(expr\) maps an abstract state to its corresponding concrete expression.
\end{itemize}

Mapping these concepts to the two domains, we get:

\paragraph{Explicit Domain.} Tracks a set of variables explicitly:
\begin{itemize}[noitemsep,nosep]
    \item $S$: A variable assignment of each \emph{tracked} variable to a value of its domain, extended with top (arbitrary value) and bottom (no assignment possible) elements. 
    \item $\top \in S$: No specific value is assigned to any of the tracked variables.
    \item $\bot \in S$: No assignment is possible to the tracked variables.
    \item $\sqsubseteq\ \subseteq S \times S$: $(s_1 \in S) \sqsubseteq (s_2 \in S) \iff (s_1 = s_2) \vee (s_1 = \bot) \vee (s_2 = \top)$
    \item ${expr}$: The conjunction of the equality expressions for each tracked variable and their value
\end{itemize}
In control-flow automata (CFA), locations are always explicitly tracked to maintain a one-to-many mapping between locations and abstract states. To mitigate state explosion, a \emph{maxenum} parameter limits value enumeration per step. For example, when a variable takes nondeterministic 32-bit values, the explicit domain keeps it at \(\top\) rather than enumerating all possibilities, enabling efficient safety checks. This makes the CEGAR loop potentially incomplete, but we can detect a non-progressing refinement cycle, and stop the analysis. 

\paragraph{Predicate Domain.} Tracks a set of boolean predicates, such that $S$ is a Boolean combination of first-order logic (FOL) predicates. $\top \in S$ is \emph{True} and $\bot \in S$ is \emph{False}. The partial order $\sqsubseteq\ \subseteq S \times S$ is the logical implication (i.e., $(s_1 \in S) \sqsubseteq (s_2 \in S) \iff (s_1 \implies s_2)$). ${expr}$ is the conjunction of the predicates.

A special version of the predicate domain is Cartesian predicate abstraction~\cite{predcart}. Here, only conjunction of ponated and negated FOL predicates are allowed in the states, as opposed to arbitrary boolean combinations of predicates.

\subsubsection{Refinement Algorithm}

Refinement either incorporates predicates from the initial precision or extracts them while refuting infeasible abstract counterexamples, gradually guiding the abstraction toward sufficient precision for verification. Two main strategies exist in \textsc{Theta}:
\begin{itemize}[noitemsep,nosep]
    \item \textbf{Single-counterexample refinement:} Refines precision based on one counterexample at a time, using binary~\cite{craiginterpolant} or sequence interpolation~\cite{thetajar}.
    \item \textbf{Multi-counterexample refinement:} Uses all counterexamples from the ARG for a combined refinement, using tree interpolation~\cite{treeinterpolant}.
\end{itemize}

The refinement strategies, if the trace is found to be spurious, return a \emph{refutation}, containing the cause of the contradiction on the path. This information is used to update the abstraction precision.

\subsection{Bounded, Property-Directed and Decision-Diagram-Based Techniques}
The second major family of techniques is implemented in the \textsc{EmergenTheta} configuration~\cite{emergentheta}. This includes bounded model checking (BMC), $k$-induction, interpolation-based model checking (IMC), property-directed reachability (PDR/IC3), and (generalized) saturation. These algorithms are defined on the \emph{symbolic transition system} (STS) formalism, which describe safety (reachability) problems using 3 SMT formulas ($I,T,P$ for \emph{initial states}, \emph{transition relation} and \emph{safety property}, respectively). For example, the STS $I: x = 0$, $T: x' = x + 1$, $P: x < 5$ describes a state space where the $x$ variable has the initial value $0$, $x$ gets incremented by $1$ when a transition fires and the safety property states that in all safe states $x$ is lower than $5$. The CFAs created from CHC problems are transformed into STS by encoding the control location as a data variable and transforming the operations to characteristic functions. \textsc{EmergenTheta} also supports chainable STS-to-STS transformations that can be used to encode enhanced analyses into STS problems and thus allow for \emph{reversed exploration}, \emph{implicit predicate abstraction} \cite{implicitabstr} and \emph{liveness checking} \cite{l2s}. Our CFA-to-STS transformation currently only supports single procedure CFAs, which means that the algorithms of the \textsc{EmergenTheta} configuration can only be used with the \emph{forward} \cite{bottomsup} CHC-to-CFA mapping.


\subsubsection{Bounded Techniques}

The bounded techniques of \textsc{EmergenTheta} focus on finding bugs or constructing inductive invariants within bounded or unrolling-based search spaces. Bounded model checking \cite{bmc} (with a loop-free check to detect finite state spaces) checks for violations of the safety property up to a finite iteratively incremented bound. $K$-induction \cite{kind} and interpolation-based model checking \cite{imc} can complement BMC with additional checks that attempt to prove that the model is safe. In case of a safe verdict, IMC can provide an (overapproximating) inductive invariant that can be used to construct a model for the original CHC problem.

\subsubsection{Property-Directed Reachability}

Property-directed reachability \cite{pdr} analyzes the model incrementally through the proof of several lemmas, which eventually form an inductive invariant proving the system correct or direct the search towards a counterexample. Initially designed for hardware model-checking, this algorithm can efficiently handle Boolean variables, but requires abstraction to reason about integer domains. Our current implementation of PDR uses STS as input and does not exploit the structural information present in CFAs. We also have an experimental implementation of the "two-dimensional" IC3 algorithm described in \cite{lange2020ic3}, which handles structural knowledge present in CFAs explicitly, but this configuration is not yet stable enough to be included in the portfolio for CHCs.

\subsubsection{Decision-Diagram-Based Techniques}

\begin{figure}[H]
    \centering
    \includegraphics[width=0.8\textwidth]{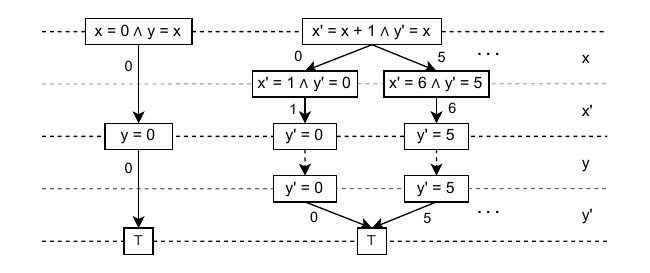}
    \caption{Substitution diagram characterizing the initial states $I: x = 0 \land y = x$ (left) and 2k-level substitution diagram characterizing the transition relation $T: x' = x + 1 \land y' = x$ (right). Each level has an associated variable, nodes correspond to SMT formulas, while the edges represent substituting the variable of the current level with a literal value in the SMT formula of the source node. We omit edges leading to the terminal $\bot$ node. The diagram on the right has infinitely many edges leaving the top node, we only display $0$ and $5$ for readability.}
    \label{fig:mdd}
\end{figure}

Decision diagrams \cite{vincephd} offer a compact way to represent sets of vectors. Substitution diagrams \cite{substdiag24} (see Fig. \ref{fig:mdd}) allow us to build multi-valued decision diagrams (MDD) from the SMT formulas of the STS models in a top-down manner, without the need to explicitly enumerate all possible valuations beforehand. For an STS model with $k$ variables, substitution diagrams with $k$ levels are used to represent the initial states and the safety property, while a diagram with $2k$ levels is used to characterize the transition relation. \emph{Generalized saturation} (GSAT) \cite{vincephd} can be used to enumerate the state space characterized by decision diagrams in a manner that decomposes exploration into smaller ones on submodels exploiting the locality of transitions. The saturation algorithm constructs a decision diagram describing all reachable states of the system, which can be used as a basis for model generation of the original CHC problem. A current limitation of this approach is that it can only handle finite state spaces, which we plan to alleviate via abstraction. We currently cannot wrap the saturation algorithm in a CEGAR loop with implicit predicate abstraction, because it does not yet yield a diagnostic trace for unsafe models.

\subsection{CHC-Solvers as Backend}
Theta also includes an experimental configuration known as \textsc{Thorn}, which integrates external CHC solvers to discharge verification conditions or analyze the CHC system directly. Because we feel this would be unfair to the external CHC solvers, who are participants of CHC-COMP themselves, we do not use the \textsc{Thorn} configuration at CHC-COMP.

\subsection{Portfolio Strategy}
\label{portfolio}

For CHC-COMP, Theta employs a sequential-portfolio-based approach, combining the CEGAR and bounded verification techniques. This allows us to balance thoroughness and efficiency, applying lightweight methods where appropriate while falling back on abstraction refinement when deeper reasoning is needed. This design enables Theta to handle a broad spectrum of CHC tasks, from shallow counterexample detection to complex proofs requiring inductive invariants.

At CHC-COMP'25, we use a sequential combination of the following configurations: 

\begin{enumerate}[noitemsep,nosep]
    \item Bounded model checking (BMC),
    \item K-induction (kIND),
    \item Interpolation-based model checking (IMC),
    \item Generalized saturation (GSAT),
    \item Boolean predicate abstraction with backwards-binary interpolation (BOOL),
    \item Cartesian predicate abstraction with backwards-binary interpolation (CART),
    \item Explicit-value abstraction with sequential interpolation (EXPL),
    \item Fallback configuration with our SV-COMP'25 portfolio~\cite{theta-svcomp}.
\end{enumerate}

The exact order of the first 7 configurations depend on the arithmetic of the task (based on empirical experience), and the first 3 may only be used with linear clauses (as these only support the \emph{backward} transformation~\cite{bottomsup}). For all configurations, we use either \textsc{z3}~\cite{z3} 4.3.0 (which is outdated, but interpolates much better than the newer versions of \textsc{z3}), or \textsc{cvc5}~\cite{cvc5} 1.0.8.

While all configurations feature a timeout value, we also handle other sources of premature completion of incomplete configurations, such as SMT solver issues, or runtime exceptions. 

Even though there are many potential tweaks we could make to this portfolio to marginally enhance its performance, such as changing SMT solvers dynamically, we do not believe these provide enough advantages to overcome the disadvantage of a more fragmented, and harder to maintain portfolio. 

\subsection{Model Generation}
\label{modelgen}

While all configurations mentioned in \autoref{portfolio} can prove safety, not all of them provide an easy-to-interpret proof that can be converted into a model for the original CHC problem. However, a few (namely: EXPL, BOOL, CART, IMC and GSAT) already provide an overabstraction of the state space as a formula, and thus, can be transformed into a model easily. 

Because of our pre-processing step transforming CHCs to CFA, we can rely on a known mapping from locations to CHC predicates. Given this knowledge, we construct a formula for all variables in the program for each location, then -- using our knowledge of the original predicate -- extract which variables are of interest, and existentially quantify the rest.

As an example, if a CFA has the variables $\{x, y, z\}$, but a predicate only takes a single argument $x$: $\mathit{inv}(x)$, then a formula $(\mathit{loc} = \mathit{inv}) \Longrightarrow x = y + 2 \land y = z + 1$ will be transformed into the model 

\begin{equation*}
\forall x : inv(x) \iff \exists y,z : x = y + 2 \land y = z + 1
\end{equation*}

While this could be further simplified in some cases, we do not (yet) perform any simplifications on our models. Furthermore, we do not yet output a trace or counterexample for the \texttt{unsat} case.

\section{Software Architecture}
\label{architecture}

\textsc{Theta} is built on a modular, layered architecture that separates concerns across four key layers: \emph{frontends}, \emph{formalisms}, \emph{algorithms}, and \emph{SMT solvers}. This architecture enables flexibility, reuse, and easy experimentation with different verification techniques and input formats.

\subsection{Frontends} The frontend layer is responsible for parsing and preprocessing the input model. For CHC solving, we use our own custom parser for the \textsc{Smt-Lib v2} format, designed specifically to support \texttt{Horn} clauses. This parser converts logical formulae into an internal intermediate representation that supports programmatic analysis and transformation.

\subsection{Formalisms} In \textsc{Theta}, all verification algorithms operate on an internal formalism. For CHC programs, this is an \emph{extended control-flow automaton} (XCFA), which augments classical CFAs with additional features such as procedure calls and threads. When using certain techniques such as bounded- or saturation-based analyses, the XCFA is further translated into a symbolic transition system that enables efficient state space exploration using only logical formulae.

\subsection{Algorithms} The architecture supports a range of verification algorithms. The algorithm layer operates independently of the input language and solver, relying only on the abstract view provided by the formalism layer. This design enables algorithm reuse across frontends and verification domains.

\subsection{SMT Solvers} The solver layer provides decision procedures for logical queries during verification. \textsc{Theta} can use \texttt{Z3}~\cite{z3} via its Java API, any solver compliant with the \textsc{Smt-Lib v2} standard~\cite{smtlibv2}, or a wide selection of up-to-date solvers through \emph{JavaSMT}~\cite{javasmt} -- an abstraction layer over modern SMT solvers such as Z3, CVC4, CVC5, MathSAT, SMTInterpol, and Boolector.

\vspace{1em}

Overall, the layered design of \textsc{Theta} allows each component to evolve independently and be extended with minimal effort, which has enabled rapid development of CHC solving capabilities on top of the existing software verification infrastructure.

\section{Strengths and Weaknesses of the Approach}
\label{strengths}

While \textsc{Theta} has been a participant at CHC-COMP for 3 years now, the main focus of it has not been to solve CHCs, but to show how a general-purpose verification framework fares compared to dedicated CHC solvers with a light-weight pre-processing step~\cite{bottomsup}.

This lack of dedication led to an oversight when assembling the release for CHC-COMP'25. We omitted a very important flag that allows \textsc{Theta} to call itself as a subprocess, thus allowing precise time- and memory-usage control. Without this option, the first configuration of the sequential portfolios were running without time limits, and thus, the whole portfolio's performance was severely hurt. We often use a lightweight (and therefore quick) but quite underpowered configuration for starting our portfolio in the case of linear tasks, the supposed strength of \textsc{Theta}. As we did not uncover this problem in time for submitting the final version, unfortunately, we finished quite behind other competitors in most of the categories. The official results (including not-inconsistent \texttt{sat} and \texttt{unsat} results, as well as ranks) can be seen in the top rows of \autoref{table:results}.

Furthermore, we accepted some unsound results due to a variable naming bug, and a buggy loop unrolling pass. This led to some tasks where our verdict differs from the rest of the competition's participants, and therefore, these tasks were deemed inconsistent, and removed from the competition. We hope that this is possible to fix retroactively, because now we know that our results were indeed wrong.

Because the fix was easy once we figured out these problems, we have since published a patch of \textsc{Theta}\footnote{https://github.com/ftsrg/theta/releases/tag/v6.15.3} that corrects this error. We would like to discuss the strength and weaknesses of the fixed version in the rest of this section, because we believe that it reflects the state of \textsc{Theta}'s CHC solving capabilities much better than the official results.

To this end, we aim to answer the following research questions:

\begin{description}
\item[RQ1\label{rq1}] How does the fixed version of our CHC-COMP configuration behave on the benchmarks of CHC-COMP'25 in terms of \emph{soundness} and \emph{performance}?
\item[RQ2\label{rq2}] Are all configurations in the portfolio capable of solving some unique tasks? 
\item[RQ3\label{rq3}] Is the performance of he portfolio better than any single configuration, and close to a theoretically optimal solution?
\end{description}

To answer \ref{rq1}, we used the same benchmarks as the official competition\footnote{https://github.com/chc-comp/chc-comp25-benchmarks/commit/ceec2f7740478cc9f114aa87fc54fc167738acdf} to run the experiments on our fixed tool again. The results can be seen in the bottom rows of \autoref{table:results}, alongside a hypothetical rank \textsc{Theta} would have achieved, had we submitted this fixed version of the tool. 
There were no results that contradict the published verdicts of the benchmark set, other than some, where the published verdict corresponds to the result produced by the submitted version of \textsc{Theta}. We believe these are erroneous, and have flagged this for the competition organizers.

\begin{table}[t]
\centering
\scriptsize
\begin{tabular}{lrC{1.75cm}C{1.75cm}C{1.75cm}C{2cm}C{1.75cm}C{1.75cm}}
\toprule
 &  & LIA & LIA-Lin & LIA-Arrays & LIA-Lin-Arrays & LRA-Lin & BV \\
 \midrule
\multirow{3}{*}{\rot{comp}}     & sat   & 52    & 114   & 400   & 45            & 73 & 49 \\
                                & unsat & 140   & 376   & 8     & 4             & 18 & 123 \\
                                & rank  & 5     & 8     & 5     & 4             & 3  & 2 \\
 \midrule
\multirow{3}{*}{\rot{fixed}}    & sat   & 48    & 585   & 440   & \textbf{63}   & 76 & 42   \\
                                & unsat & 136   & 402   & 12    & 18            & 16 & 126  \\
                                & rank  & 5     & 5     & 4     & 1             & 3  & 2    \\
\bottomrule
\end{tabular}
\caption{Results of \textsc{Theta} at the competition (comp) and with the configuration fix (fixed).}
\label{table:results}
\end{table}

In order to answer \ref{rq2}, we counted the number of tasks a configuration solved in a category (keeping in mind, that a prior successful configuration disables a later configuration, so we naturally expect diminished results the later a configuration is). These results are shown in \autoref{table:statistic}. Note that these include the inconsistent tasks as well, while \autoref{table:results} does not.

Furthermore, for the linear category -- them being our focus -- we measured each configuration on its own, and also calculated a "virtual best" configuration (taking the best performing configuration for each successfully solved task, therefore representing the result of a theoretical optimal algorithm-selection) to answer \ref{rq3}. These results are shown in \autoref{fig:quantileplots}.

\begin{table}[t]
\centering
\scriptsize
\begin{tabular}{@{}rC{1.75cm}C{1.75cm}C{1.75cm}C{2cm}C{1.75cm}C{1.75cm}@{}}
\toprule
        & LIA  & LIA-Lin   & LIA-Arrays    & LIA-Lin-Arrays    & LRA-Lin   & BV \\ \midrule
BMC     &      & 450       &               & 26                & 88        & 17 \\
kIND    &      & 35        &               & 30                & 1         & 186\\
IMC     &      & 235       &               & 22                &           &    \\
GSAT    &      & 4         &               &                   &           &    \\
BOOL    & 93   & 21        & 395           & 25                & 9         & 13 \\
CART    & 35   & 163       & 23            & 25                & 1         & 1  \\
EXPL    & 58   & 91        & 8             & 22                &           & 3  \\
SVCOMP  &      &           & 26            &                   &           &    \\ \bottomrule
\end{tabular}
\caption{Solved tasks per configuration per category.}
\label{table:statistic}
\end{table}

\begin{figure}
    \centering

    \begin{subfigure}{0.5\textwidth}
    \input{quantile-LIA}
    \caption{LIA-Lin}
    \label{fig:lia-lin}
    \end{subfigure}%
    \begin{subfigure}{0.5\textwidth}
    \input{quantile-LIA-Arrays}
    \caption{LIA-Lin-Arrays}
    \label{fig:lia-lin-arrays}
    \end{subfigure}
    
    \begin{subfigure}{0.5\textwidth}
    \input{quantile-BV}
    \caption{BV}
    \label{fig:bv}
    \end{subfigure}%
    \begin{subfigure}{0.5\textwidth}
    \input{quantile-LRA}
    \caption{LRA-Lin}
    \label{fig:lra-lin}
    \end{subfigure}
    
    \caption{Quantile plots of configurations in the linear categories}
    \label{fig:quantileplots}
\end{figure}

\subsection{Analysis of the Results}

From \autoref{table:results}, we can see there are some categories where the fixed version underperforms the competition configuration (namely: \texttt{LIA} for both, \texttt{LRA-Lin} for \texttt{unsat}, and \texttt{BV} for \texttt{sat} results). These do not change the ranking.

We believe these are due to the variable naming and unrolling fixes, and previously, we just happened to find the correct result, but by an erroneous reasoning process.

However, for the majority of categories and results, our fixed solution outperforms the submitted version greatly. While in most cases this does not affect the rank, in 2 categories, it improves it: 

\begin{enumerate}
    \item \texttt{LIA-Arrays}: from 5th place to 4th,
    \item \texttt{LIA-Lin-Arrays}: from 4th place to 1st.
\end{enumerate}

Additionally, the bold \textbf{63} number in the \texttt{LIA-Lin-Arrays} category's \texttt{sat} row shows the highest number of solved tasks among the competition participants. Therefore (at least on the CHC-COMP'25 benchmark set), our tool is the most successful at proving satisfiability of a CHC system with arrays, and only linear clauses.

\begin{description}
    \item[\ref{rq1}:] There are no unsound results (when compared to the other participants when they are in agreement), and in two categories, our ranking would improve.
\end{description}

\autoref{table:statistic} shows the number of solved tasks per configuration, per category. There are some categories, where a configuration produced no results, but these may be caused by them being ordered at the end of the sequential portfolio, and thus, an earlier successful configuration would take their chance away from solving the task at hand. However, throughout all categories, every single configuration we feature in our portfolio was successful in at least some of the task. Furthermore, based on further experiments that resulted in \autoref{fig:quantileplots}, we know that no single configurations is completely \emph{shadowed} by another, as there were tasks that were uniquely solved by a single configuration.    

\begin{description}
    \item[\ref{rq2}:] All configurations in the portfolio solved some tasks in some categories.
\end{description}

In \autoref{fig:quantileplots}, we show the portfolio results in \textbf{\textcolor{ftsrg@AccentRed}{bold red}} and the "virtual best" configuration (taking the quickest correct result from any of the single configuratoins) in \textcolor{ftsrg@AccentBlue}{dotted blue}. We also show the best 4 configurations in each category. 

In most cases, the portfolio follows the virtual best selection relatively closely. In \autoref{fig:lia-lin-arrays}, it is almost the same curve, in \autoref{fig:lia-lin} and \autoref{fig:lra-lin} the portfolio is slower but solves almost the same number of tasks, but in \autoref{fig:bv}, the virtual best is much better than the portfolio. Also, k-induction solves almost the same number of tasks as the portfolio, and even quicker in most cases. In the other categories, no single configuration outperforms the portfolio at any point in the runtime meaningfully.

\begin{description}
    \item[\ref{rq3}:] In most categories (with linear clauses), the portfolio closely follows the virtual best selection and outperforms all single configurations, but with bitvector arithmetic, the portfolio is far from the virtual best, and k-induction is often better. 
\end{description}

These results indicate that the current portfolio strategy is suboptimal for the \texttt{BV} category, and that further tuning or diversification of configurations is necessary to improve performance

\subsection{Threats to Validity}
\label{sec:threats-to-validity}

The following factors may influence the validity of our experiments.

\emph{Internal validity.}
We used BenchExec~\cite{benchexec} to ensure accuracy. We ran our experiments on virtual machines in the cloud computing platform of our university. External factors such as loads on other virtual machines of the host and shared resources may have influenced the results.

\emph{External validity.}
The CHC-COMP benchmark suite is considered the standard for academic benchmarking of CHC solving algorithms and tools. Still, evaluation results might not generalize well to other sources of problems. 

\emph{Construct validity.}
The metrics of the evaluation were carefully chosen to accurately describe the performance of our algorithm, while not using misleading statistics such as memory usage, given our tool is running in a managed environment (JVM). We concentrate on statistics that meaningfully impact a potential user of our tool: the number of solved tasks, and the time it takes to produce a solution.

\section{Tool Setup and Configuration}
\label{toolsetup}

\textsc{Theta} is a modular and highly configurable framework~\cite{thetajar}, with support for multiple input languages, algorithms, and solvers. For CHC-based verification tasks, we recommend using the following invocation: \texttt{./chc <input>}. This runs our portfolio-based approach, which chooses a suitable sequence of configurations based on the input file automatically. 

\section{Software Project and Data Availability}
\label{softproj}

\textsc{Theta} is developed and maintained by the Critical Systems Research Group at the Budapest University of Technology and Economics. The framework is available open-source on GitHub\footnote{https://github.com/ftsrg/theta} under the Apache 2.0 license. The version used for the experiments in this paper is available at~\cite{zenodo-chc-version}.

\bibliographystyle{splncs04}
\bibliography{chc}

\end{document}

%% file: figures/cegarloop.tex
\begin{figure}
\begin{minipage}[b]{0.66\textwidth}
\begin{figure}[H]
\centering
\begin{tikzpicture}[scale=1, every node/.style={transform shape},x={(0.75cm,0cm),y={(0cm,0.75cm)}}]
    \tikzset{vertex/.style = {shape=ellipse,draw,minimum size=1.5em}}
    \node[vertex,draw=none] (Initprec) at  (0,0) {Initial precision};
    \node[vertex,shape=rectangle] (Abs) at  (0,-1.5) {Abstractor};
    \node[vertex,shape=rectangle] (Ref) at  (9,-1.5) {Refiner};
    \node[vertex] (ARG) at  (4.5,-1.5) {ARG};
    \node[vertex] (Safe) at  (0,-3) {Safe};
    \node[vertex] (Unsafe) at  (9,-3) {Unsafe};
    \tikzset{edge/.style = {->,> = latex',color=black}}
    \draw[edge] (Initprec) to (Abs);
    \draw[edge] (Abs) to (Safe);
    \draw[edge] (Ref) to (Unsafe);
    \draw[edge] (Abs) to (2,-0.5) to node [above] {Abstract counterexample} (7,-0.5) to (Ref);
    \draw[edge] (Ref) to (7,-2.5) to node [below] {Refined precision} (2,-2.5) to (Abs);
    \draw[edge,dashed] (Abs) to node [above] {Expand} (ARG);
    \draw[edge,dashed] (Ref) to node [above] {Prune} (ARG);
\end{tikzpicture}
\caption{The CEGAR loop}
\label{fig:cegarloop}
\end{figure}
\end{minipage}
\begin{minipage}[b]{0.31\textwidth}
\begin{figure}[H]
    \centering
    \includegraphics[width=0.8\textwidth]{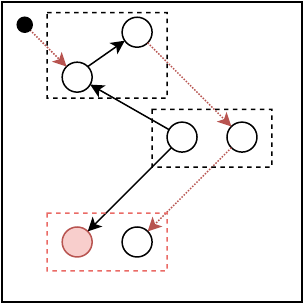}
    \caption{An ARG}
    \label{fig:arg}
\end{figure}
\end{minipage}
\end{figure}

%% file: quantile-LIA.tex
\pgfplotstableread[col sep=comma]{quantile-LIA.csv}\datatable
\begin{tikzpicture}[scale=1, every node/.style={transform shape}]

\tikzset{mymark/.style={
        decoration={
            markings,
            mark= between positions 0 and 1 step 5mm with
                {
                \pgfuseplotmark{#1};
            },
        },
        postaction={decorate}
    }
}

\begin{axis}[
width=0.9\columnwidth,
height=0.5\columnwidth,
xtick distance={250},
x tick label style={font=\small, rotate=0, anchor=north},
y tick label style={font=\small, rotate=0, anchor=east},
legend style={at={(0.5,1)},anchor=south,font=\tiny,legend columns=3},
ylabel={\small{Time (s)}},
xlabel={\small{Tasks}},
ymode=log,
log ticks with fixed point,
grid=both,
grid style={gray!20},
xmin=0,xmax=1050,
ymin=1,ymax=2000,
every axis plot/.append style={ultra thick}]

\addplot [ftsrg@AccentGreen, line width=1pt, /pgf/number format/read comma as period=true] table [x expr=\coordindex, y={EXPL}]{\datatable};
\addlegendentry{EXPL}

\addplot [ftsrg@AccentOrange, line width=1pt, /pgf/number format/read comma as period=true] table [x expr=\coordindex, y={KIND}]{\datatable};
\addlegendentry{KIND}

\addplot [ftsrg@AccentPurple, line width=1pt, /pgf/number format/read comma as period=true] table [x expr=\coordindex, y={CART}]{\datatable};
\addlegendentry{CART}

\addplot [ftsrg@AccentLightBlue, line width=1pt, /pgf/number format/read comma as period=true] table [x expr=\coordindex, y={BOOL}]{\datatable};
\addlegendentry{BOOL}

\addplot [ftsrg@AccentRed, line width=2pt, /pgf/number format/read comma as period=true] table [x expr=\coordindex, y={theta-portfolio}]{\datatable};
\addlegendentry{Portfolio}

\addplot [ftsrg@AccentBlue, densely dotted, line width=1pt, /pgf/number format/read comma as period=true ] table [x expr=\coordindex, y={virtual-best}]{\datatable};
\addlegendentry{Virtual Best}

\end{axis}
\end{tikzpicture}

%% file: quantile-LIA-Arrays.tex
\pgfplotstableread[col sep=comma]{quantile-LIA-Arrays.csv}\datatable
\begin{tikzpicture}[scale=1, every node/.style={transform shape}]

\tikzset{mymark/.style={
        decoration={
            markings,
            mark= between positions 0 and 1 step 5mm with
                {
                \pgfuseplotmark{#1};
            },
        },
        postaction={decorate}
    }
}

\begin{axis}[
width=0.9\columnwidth,
height=0.5\columnwidth,
xtick distance={25},
x tick label style={font=\small, rotate=0, anchor=north},
y tick label style={font=\small, rotate=0, anchor=east},
legend style={at={(0.5,1)},anchor=south,font=\tiny,legend columns=3},
ylabel={\small{Time (s)}},
xlabel={\small{Tasks}},
ymode=log,
log ticks with fixed point,
grid=both,
grid style={gray!20},
xmin=0,xmax=100,
ymin=1,ymax=2000,
every axis plot/.append style={ultra thick}]

\addplot [ftsrg@AccentGreen, line width=1pt, /pgf/number format/read comma as period=true] table [x expr=\coordindex, y={EXPL}]{\datatable};
\addlegendentry{EXPL}

\addplot [ftsrg@AccentOrange, line width=1pt, /pgf/number format/read comma as period=true] table [x expr=\coordindex, y={KIND}]{\datatable};
\addlegendentry{KIND}

\addplot [ftsrg@AccentPurple, line width=1pt, /pgf/number format/read comma as period=true] table [x expr=\coordindex, y={CART}]{\datatable};
\addlegendentry{CART}

\addplot [ftsrg@AccentLightBlue, line width=1pt, /pgf/number format/read comma as period=true] table [x expr=\coordindex, y={BOOL}]{\datatable};
\addlegendentry{BOOL}

\addplot [ftsrg@AccentRed, line width=2pt, /pgf/number format/read comma as period=true] table [x expr=\coordindex, y={theta-portfolio}]{\datatable};
\addlegendentry{Portfolio}

\addplot [ftsrg@AccentBlue, densely dotted, line width=1pt, /pgf/number format/read comma as period=true ] table [x expr=\coordindex, y={virtual-best}]{\datatable};
\addlegendentry{Virtual Best}

\end{axis}
\end{tikzpicture}

%% file: quantile-BV.tex
\pgfplotstableread[col sep=comma]{quantile-BV.csv}\datatable
\begin{tikzpicture}[scale=1, every node/.style={transform shape}]

\tikzset{mymark/.style={
        decoration={
            markings,
            mark= between positions 0 and 1 step 5mm with
                {
                \pgfuseplotmark{#1};
            },
        },
        postaction={decorate}
    }
}

\begin{axis}[
width=0.9\columnwidth,
height=0.5\columnwidth,
xtick distance={50},
x tick label style={font=\small, rotate=0, anchor=north},
y tick label style={font=\small, rotate=0, anchor=east},
legend style={at={(0.5,1)},anchor=south,font=\tiny,legend columns=3},
ylabel={\small{Time (s)}},
xlabel={\small{Tasks}},
ymode=log,
log ticks with fixed point,
grid=both,
grid style={gray!20},
xmin=0,xmax=200,
ymin=1,ymax=2000,
every axis plot/.append style={ultra thick}]

\addplot [ftsrg@AccentGreen, line width=1pt, /pgf/number format/read comma as period=true] table [x expr=\coordindex, y={BMC}]{\datatable};
\addlegendentry{BMC}

\addplot [ftsrg@AccentOrange, line width=1pt, /pgf/number format/read comma as period=true] table [x expr=\coordindex, y={KIND}]{\datatable};
\addlegendentry{KIND}

\addplot [ftsrg@AccentPurple, line width=1pt, /pgf/number format/read comma as period=true] table [x expr=\coordindex, y={CART}]{\datatable};
\addlegendentry{CART}

\addplot [ftsrg@AccentLightBlue, line width=1pt, /pgf/number format/read comma as period=true] table [x expr=\coordindex, y={BOOL}]{\datatable};
\addlegendentry{BOOL}

\addplot [ftsrg@AccentRed, line width=2pt, /pgf/number format/read comma as period=true] table [x expr=\coordindex, y={theta-portfolio}]{\datatable};
\addlegendentry{Portfolio}

\addplot [ftsrg@AccentBlue, densely dotted, line width=1pt, /pgf/number format/read comma as period=true ] table [x expr=\coordindex, y={virtual-best}]{\datatable};
\addlegendentry{Virtual Best}

\end{axis}
\end{tikzpicture}

%% file: quantile-LRA.tex
\pgfplotstableread[col sep=comma]{quantile-LRA.csv}\datatable
\begin{tikzpicture}[scale=1, every node/.style={transform shape}]

\tikzset{mymark/.style={
        decoration={
            markings,
            mark= between positions 0 and 1 step 5mm with
                {
                \pgfuseplotmark{#1};
            },
        },
        postaction={decorate}
    }
}

\begin{axis}[
width=0.9\columnwidth,
height=0.5\columnwidth,
xtick distance={25},
x tick label style={font=\small, rotate=0, anchor=north},
y tick label style={font=\small, rotate=0, anchor=east},
legend style={at={(0.5,1)},anchor=south,font=\tiny,legend columns=3},
ylabel={\small{Time (s)}},
xlabel={\small{Tasks}},
ymode=log,
log ticks with fixed point,
grid=both,
grid style={gray!20},
xmin=0,xmax=100,
ymin=1,ymax=2000,
every axis plot/.append style={ultra thick}]

\addplot [ftsrg@AccentGreen, line width=1pt, /pgf/number format/read comma as period=true] table [x expr=\coordindex, y={BMC}]{\datatable};
\addlegendentry{BMC}

\addplot [ftsrg@AccentOrange, line width=1pt, /pgf/number format/read comma as period=true] table [x expr=\coordindex, y={KIND}]{\datatable};
\addlegendentry{KIND}

\addplot [ftsrg@AccentPurple, line width=1pt, /pgf/number format/read comma as period=true] table [x expr=\coordindex, y={CART}]{\datatable};
\addlegendentry{CART}

\addplot [ftsrg@AccentLightBlue, line width=1pt, /pgf/number format/read comma as period=true] table [x expr=\coordindex, y={BOOL}]{\datatable};
\addlegendentry{BOOL}

\addplot [ftsrg@AccentRed, line width=2pt, /pgf/number format/read comma as period=true] table [x expr=\coordindex, y={theta-portfolio}]{\datatable};
\addlegendentry{Portfolio}

\addplot [ftsrg@AccentBlue, densely dotted, line width=1pt, /pgf/number format/read comma as period=true ] table [x expr=\coordindex, y={virtual-best}]{\datatable};
\addlegendentry{Virtual Best}

\end{axis}
\end{tikzpicture}